\documentclass[oribibl]{llncs}
\usepackage{graphicx, graphics,bm, natbib, url}
\usepackage{verbatim}
\usepackage[english]{babel}
\usepackage{multicol}
\usepackage[latin1]{inputenc}
\usepackage{times}
\usepackage[T1]{fontenc}
\usepackage{amsmath}
\usepackage{color}
\usepackage{float}
\usepackage{caption}
\usepackage{setspace}
\usepackage{multirow}
\usepackage{bbm}
\usepackage[T1]{fontenc}
\usepackage{authblk}
\usepackage{multirow}
\usepackage{amssymb}

\begin{document}
\title{\large \bf Two-Phase Data Synthesis for Income: An Application to the NHIS}
\author{Kevin Ros, Henrik Olsson and Jingchen Hu\footnote{This work is published in the e-proceedings of Privacy in Statistical Databases 2020 and presented at the conference.}}
\institute{Vassar College, Poughkeepsie, NY 12604, USA}
\maketitle

\begin{abstract}
We propose a two-phase synthesis process for synthesizing income, a sensitive variable which is usually highly-skewed and has a number of reported zeros. We consider two forms of a continuous income variable: a binary form, which is modeled and synthesized in phase 1; and a non-negative continuous form, which is modeled and synthesized in phase 2. Bayesian synthesis models are proposed for the two-phase synthesis process, and other synthesis models are readily implementable. We demonstrate our methods with applications to a sample from the National Health Interview Survey (NHIS). Utility and risk profiles of generated synthetic datasets are evaluated and compared to results from a single-phase synthesis process. 
\end{abstract}

\section{Introduction}
\label{intro}

It is impossible to overstate the importance of data in today's world. Nearly every decision made by corporations and governments is based off conclusions drawn from data, in one way or another. However, many datasets contain detailed information at the individual level, which leads to the possibility of malicious intruders obtaining confidential information about individuals in the datasets, commonly known as disclosure risk~\citep{DomingoTorra2004JCAM, SkinnerShlomo2008JASA, hu2018bayesian}. Not only can this harm the individuals but it may also lead to the disclosure of legally protected information, such as medical records \citep{Sweeney1997JLME}. 

One approach to mitigating such issues, especially for microdata (i.e. record-level data on individuals or business establishments), is to simulate and release synthetic data \citep{Rubin1993synthetic, Little1993synthetic, RaghuReiterRubin2003JOS, Drechsler2011book}. Data holders, such as statistical agencies and organizations, develop statistical models on the original, confidential dataset, and simulate synthetic records based on the estimated models. These synthetic records are released to the public in place of the original, confidential records.

Income information has long been the subject of data synthesis, due to its sensitivity and prevalence. \citet{kennickell1997multiple} attempts to reduce disclosure risks from the Survey of Consumer Finances (containing families' balance sheets) by various data synthesis techniques. \citet{abowd2001disclosure} investigate disclosure risk regarding longitudinal income data. Income has also been synthesized using classification and regression trees methods (CART), as in \citet{reiter2005using}. Additionally, non-parametric and parametric synthesizing techniques have been used to circumvent some issues of income top-coding 
\citep{an2007multiple}. More recently, \citet{respp_arxiv} propose vector-weighted synthesis methods with applications to the Consumer Expenditure Survey (CE) income data. \citet{SavitskyWilliamsHu2020ppm} develop synthesis models under differential privacy \citep{Dwork:2006:CNS:2180286.2180305} with applications to the CE income. \citet{reweight_arxiv} evaluated differentially private synthetic microdata for the sensitive salary information in the Survey of Doctorate Recipients.


The income variable is particularly sensitive due to its uniqueness and potential for outliers. In certain datasets, the income variable is highly-skewed, and may contain a significant number of zeros (individuals with no income) along with non-zeros (individuals with income). These features can cause certain synthesis models to lose their effectiveness, resulting in synthetic data with little usefulness. 

To tackle these challenges, we propose a two-phase synthesis process, where we consider two forms of a continuous income variable: a binary form, indicating zero income and non-zero income; and a non-negative continuous form, containing all records with a non-zero income value. In a sequential manner, we first model and synthesize the binary form of income in phase 1. Next in phase 2, the continuous form of income is modeled and synthesized, only for the records receiving a non-zero income from phase 1. Similar approaches have been proposed and implemented for missing data imputation of income \citep{raghunathan2001multivariate}. 

Specifically, we consider Bayesian synthesis models for the two-phase synthesis process: a Bayesian logistic regression synthesis model in phase 1, and a Bayesian linear regression synthesis model in phase 2. We apply our two-phase Bayesian synthesis models to a data sample from the National Health Interview Survey (NHIS). For comparison, a single-phase Bayesian synthesis model is implemented. Our results show that the two-phase synthesis approach produces synthetic data with much higher utility, at the cost of slightly higher identification and attribute disclosure risks, compared to the single-phase synthesis approach. Other synthesis models can be readily implemented in our proposed two-phase synthesis process.





We now introduce the details of the NHIS data sample in our applications.

\subsection{The NHIS data sample}
\label{intro:data}



The NHIS is an annual survey, collecting individual-level survey data for numerous research purposes. The collected, self-reported data includes extensive information on the demographic, socioeconomic, and health experiences of individuals living in the U.S. In our applications, we select on a random sample of 5000 observations from the 2018 NHIS. 

\begin{table}
\begin{center}
\begin{tabular}{p{2.55cm}p{2cm}p{7.25cm}}
 \hline
Variable & Type & Description \\
 \hline
Income & Continuous & 0 - 149,000 \\
Age & Continuous & 18 - 85 \\
Sex & Binary & 1 = male, 2 = female\\
Race & Categorical & 1 = White, 2 = African-American, 3 = American Indian, 4 = Asian, 5 = other races \\
Education & Categorical & 1 = 4 years of high school or less, 2 = 1 - 4 years of college,  3 = 5+ years of college \\
HoursWorked & Continuous & 1 - 95+ \\
HealthInsurance & Binary & 1 = no, has coverage, 2 = yes, has no coverage \\
HomeOwnership & Categorical & 1 = own, 2 = rent, 3 = other \\
 \hline
\end{tabular}
\vspace{1mm}
\caption{Variables in the NHIS data sample.}
\label{tab:variable}
\end{center}
\end{table}

Our selected NHIS data sample is composed of eight variables: Income, Age, Sex, Race, Education, HoursWorked, HealthInsurance, and HomeOwnership. Among these variables, Income, Age, and HoursWorked are continuous variables; Sex and HealthInsurance are binary variables; and Race, Education, and HomeOwnership are categorical variables. Table~\ref{tab:variable} presents the details of the variables, including variable type and description. Our data cleaning process is included in the Appendix.




Among the eight variables in the NHIS data sample, the continuous Income variable is deemed sensitive. It contains 4\% records with reported zero income value, and it is top-coded at \$149,000. Our goal is to synthesize this Income variable to provide privacy protection. The remaining variables are used as predictors in our proposed Bayesian synthesis models, resulting in partially synthetic data, where a subset of variables is synthesized \citep{Little1993synthetic}. For illustration purpose, we treat this publicly available sample as the confidential, non-topcoded data.

The remainder of this paper is outlined as follows. In Section~\ref{methods}, we describe the proposed two-phase synthesis process with Bayesian synthesis models, illustrated with synthesizing the income variable in the NHIS data sample. We present the utility and risk evaluation measures and results of our proposed two-phase synthesis models applied to the NHIS data sample in Section~\ref{results}, with comparison to the results to those from a single-phase synthesis model. We end with a few concluding remarks in Section~\ref{discussion}.

\section{A Two-Phase Synthesis Process}
\label{methods}

\begin{figure}[h]
    \centering
    \includegraphics[scale=0.3]{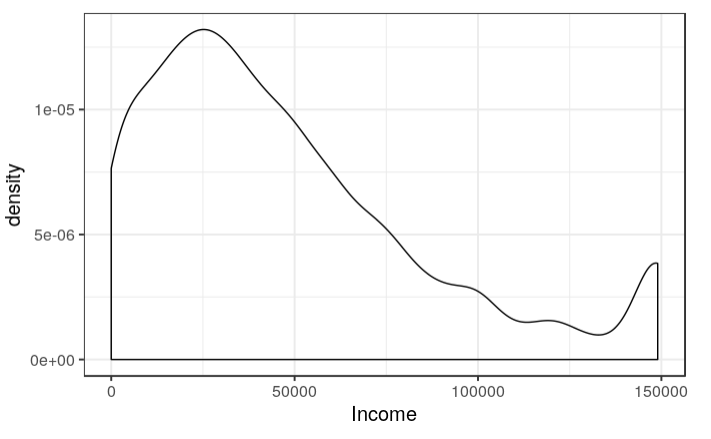}
    \caption{Density plot of the original income values.}
    \label{fig:income_orig}
\end{figure}

Our NHIS data sample contains 5000 individuals, and about 4\% of these records report a zero income value (see Figure \ref{fig:income_orig} for a density plot of the income distribution). To synthesize a highly-skewed, non-negative continuous variable with a number of 0s, a commonly used approach is a Bayesian linear regression synthesis model with appropriate data transformation. For example, we can add a small value to every observation and take the logarithm, then use a Bayesian regression synthesis model utilizing a number of predictors. 


While such an approach is straightforward, it might not be able to create synthetic income values of exactly 0, due to the data transformation step. These could have a severe negative impact on analysis such as ``percentage of zero value income": while it is 4\% in the original, confidential dataset, it is 0\% in the synthetic data. Moreover, as we will show in Section \ref{results:utility}, this approach could result in bias of various of estimates of interest, such as the mean, extreme quantiles, and regression coefficients.

We therefore propose a two-phase synthesis process, where we first create two forms of the sensitive income variable: binary (zeros vs non-zeros, labeled as ``Income-B") and continuous (positive continuous values, labeled as ``Income-C"). See Table~\ref{tab:variable_income} for details.

\begin{table}
\begin{center}
\begin{tabular}{p{2.55cm}p{2cm}p{7.25cm}}
 \hline
Variable & Type & Description \\
 \hline
Income-B & Binary & 0 = zero income, 1 = non-zero income \\
Income-C & Continuous & 1 - 149,000 (only available for records with Income-B of 1) \\
 \hline
\end{tabular}
\vspace{1mm}
\caption{Two forms of income variable in the two-phase synthesis models.}
\label{tab:variable_income}
\end{center}
\end{table}

There are two sequential steps in our proposed two-phase synthesis process. We now illustrate our process with Bayesian synthesis models, and we note that other synthesis models, such as classification and regression trees (CART) models, can be readily implemented in the two-phase process \citep{reiter2005using, RaabNowokDibben2016JPC}. 

Our phase 1 focuses on synthesizing Income-B using an appropriate Bayesian logistic regression synthesis model. Phase 2 then focuses on synthesizing Income-C, only for records with synthesized non-zero income from phase 1, using an appropriate Bayesian linear regression synthesis model. In this way, we ensure the existence of records with income values of exactly 0 in the synthetic data. Moreover, the distributional characteristics of income given the predictors in the original dataset can be well-maintained.

We proceed to describe the two-phase Bayesian synthesis models in detail. Section \ref{methods:CatIncome} describes phase 1 of synthesizing the binary Income-B using a Bayesian logistic regression model. Section \ref{methods:ConIncome} describes phase 2 of synthesizing the continuous Income-C using a Bayesian linear regression model.

\subsection{Phase 1: synthesizing binary income}
\label{methods:CatIncome}

To synthesize binary variable Income-B for all $n = 5000$ observations, we use a Bayesian logistic regression model. Let $Y_i \in \{0, 1\}$ be the outcome of variable Income-B of record $i (i = 1, \cdots, n)$. Let $p_i$ be the probability of $Y_i = 1$. $Y_i$ then follows a Bernoulli distribution with probability $p_i$:


\begin{equation}
    Y_{i} \sim \textrm{Bernoulli}(p_{i}). 
    \label{eq:bern}
\end{equation}

The logit of $p_i$ is expressed as a linear combination of available predictors. We consider the main effect of all seven predictors: 


\begin{eqnarray}
\textrm{logit}(p_{i}) &=& \beta_{0} + \beta_{1}\textrm{Age}_{i} + \beta_{2}\textrm{Sex}_{i} + \beta_{3}\textrm{Race}_{i} + \beta_{4}\textrm{Education}_{i} + \nonumber \\
&&\beta_{5}\textrm{HoursWorked}_{i} +\beta_{6}\textrm{HealthInsurance}_{i} + \beta_{7}\textrm{HomeOwnership}_{i}. 
\label{eq:logit}
\end{eqnarray}

Just Another Gibbs Sampler (JAGS) is used for Markov chain Monte Carlo (MCMC) estimation of the logistic regression synthesis model \citep{JAGS}. To generate synthetic Income-B values for every record, we first collect posterior draws of $\bm{\beta}^{(\ell)}$ at MCMC iteration $\ell$. We next simulate $\tilde{p}_i^{(\ell)}$ from Equation (\ref{eq:logit}) given $\bm{\beta}^{(\ell)}$ and all predictor values. Finally we simulate $\tilde{Y}_i^{(\ell)}$ from Equation (\ref{eq:bern}) given $\tilde{p}_i^{(\ell)}$. This completes the synthesis for record $i$, and we do so for all $n$ records in the similar fashion to obtain a synthetic vector $\tilde{\bm{Y}}^{(\ell)}$ of Income-B. To generate $m > 1$ synthetic datasets, we repeat this process at $m$ independent MCMC iterations, so that we have $m$ synthetic vectors of Income-B: $\tilde{\bm{Y}} = (\tilde{\bm{Y}}^{(1)}, \cdots, \tilde{\bm{Y}}^{(m)})$.


\subsection{Phase 2: synthesizing continuous income}
\label{methods:ConIncome}

Let $n^*_{syn}$  denote the number of records with non-zero income in the synthetic data from phase 1 ($n^*_{syn} \leq 5000$). To synthesize continuous variable Income-C for these $n^*_{syn}$ observations, we use a Bayesian linear regression model.  Let $Z_i > 0$ be the logarithm of the outcome of variable Income-C of record $i\;(i = 1, \cdots, n^*_{syn})$. $Z_i$ follows a normal distribution with mean $\mu_i$ and standard deviation $\sigma$: 



\begin{equation}
    Z_{i}\mid\mu_{i},\sigma \sim \textrm{Normal}(\mu_{i},\sigma),
    \label{eq:normal}
\end{equation}
where $\mu_i$ is a linear combination of the main effect of all seven predictors:
\begin{eqnarray}
\mu_{i} &=&  \beta^*_{0} + \beta^*_{1}\textrm{Age}_{i} + \beta^*_{2}\textrm{Sex}_{i} + \beta^*_{3}\textrm{Race}_{i} + 
\beta^*_{4}\textrm{Education}_{i} + \nonumber \\
&& \beta^*_{5}\textrm{HoursWorked}_{i} +
 \beta^*_{6}\textrm{HealthInsurance}_{i} + \beta^*_{7}\textrm{HomeOwernship}_{i}.
\label{eq:mu}
\end{eqnarray}
The prior distribution for $\sigma$ is $ 1/\sigma ^2\sim \textrm{Gamma}(\alpha^*,\beta^*)$.


As before, we use JAGS for the MCMC estimation of the regression synthesis model. To generate a synthetic Income-C value for every record $i \;(i = 1, \cdots, n^*_{syn})$ with a non-zero Income-B value from phase 1, we first collect posterior draws of $\bm{\beta}^{*, (\ell)}$ at MCMC iteration $\ell$. We next simulate $\tilde{\mu}_i^{(\ell)}$ from Equation (\ref{eq:mu}) given $\bm{\beta}^{*, (\ell)}$ and all predictor values. Finally we simulate $\tilde{Z}_i^{(\ell)}$ from Equation (\ref{eq:normal}) given $\tilde{\mu}_i^{(\ell)}$. This completes the synthesis for record $i$, and we do so for all $n_{syn}^*$ records in the similar fashion to obtain a synthetic vector $\tilde{\bm{Z}}^{(\ell)}$ of Income-C. To generate $m > 1$ synthetic datasets, we repeat this process at $m$ independent MCMC iterations, so that we have $m$ synthetic vectors of Income-C: $\tilde{\bm{Z}} = (\tilde{\bm{Z}}^{(1)}, \cdots, \tilde{\bm{Z}}^{(m)})$, each corresponds to one of the $m$ synthetic vectors from phase 1 of Income-B, $\tilde{\bm{Y}} = (\tilde{\bm{Y}}^{(1)}, \cdots, \tilde{\bm{Y}}^{(m)})$. Note that we insert a value of 0 for any record with a zero Income-B in the synthetic Income-C vectors of $\tilde{\bm{Z}}$, so that they have the same length as $\tilde{\bm{Y}}$.

Finally, $[\tilde{\bm{Y}}^{(\ell)}; \tilde{\bm{Z}}^{(\ell)}]^T$ represents the synthetic Income-B and Income-C for the $\ell$-th synthetic dataset ($\ell = 1, \cdots, m$).

\section{Utility and Risk Evaluations: Measures and Results}
\label{results}

We apply our proposed two-phase Bayesian synthesis models to the NHIS data sample introduced in Section \ref{intro:data}. In both phases, we assume weakly informative, univariate Normal(0, 1) prior for each regression coefficient in Equation (\ref{eq:logit}) and Equation (\ref{eq:mu}), and a Gamma(1, 1) prior for precision $1 / \sigma^2$ parameter in Equation (\ref{eq:normal}). We run the MCMC chains long enough to ensure convergence of all parameters and passing all relevant MCMC diagnostics. We generate $m = 20$ synthetic datasets from the two-phasis synthesis models, each containing a synthetic vector of Income-B and a synthetic vector of Income-C.

For comparison, we synthesize $m = 20$ synthetic datasets from a single-phase synthesis model described in Section \ref{methods}. In this model, we use the same set of predictors and the same set of prior distributions as in the two-phase synthesis models.

We choose $m = 20$ to adequately explore the utility and risk profiles of the two synthesis models. In practice, $m = 1$ synthetic dataset can be simulated and released for higher privacy protection \citep{ReiterMitra2009, KleinSinha2015JPC, RaabNowokDibben2016JPC}.

We now proceed to describe the evaluation measures and results of utility and risk in Section \ref{results:utility} and Section \ref{results:risk}, respectively. All results from the two-phase synthesis models are denoted with a ``t" superscript, whereas all results from the single-phase synthesis model are denoted with an ``s" superscript.



\subsection{Utility evaluations: measures and results}
\label{results:utility}

To evaluate utility of simulated synthetic datasets, we first consider two global utility measures: the propensity score measure and the empirical CDF measure.


\vspace{2mm}
\noindent {\bf Propensity score}
The propensity score measure aims to quantify the degree of a classification algorithm distinguishing the original dataset from the synthetic dataset \citep{woo2009global, Snoke2018JRSSA}. We first merge the original and synthetic datasets, resulting in a merged dataset of $2n$ records, and add an indicator variable of a record's membership (0 for original, and 1 for synthetic). We next estimate the probability $\hat{p}_i$ of record $i$ belonging to the synthetic dataset using a classification algorithm. Finally, we compute the propensity score utility measure, $U_p$:
\begin{equation}
     U_p = \frac{1}{2n}\sum_{i=1}^{2n} \left(\hat{p}_i - \frac{1}{2}\right)^2,
     \label{eq:Up}
\end{equation}
where $2n$ is the number of records in the merged dataset. On the one hand, the smaller and closer-to-0 $U_p$ is, the higher the level of similarity between the two datasets, indicating higher utility. On the other hand, the larger and closer-to-1/4 $U_p$ is, the lower the level of similarity between the two datasets, indicating lower utility.

We use a logistic regression of Income given all available predictors to estimate the propensity scores. We calculate the propensity score measure of each of the $m = 20$ merged datasets, for the two-phase and the one-phase respectively. We report the average across the $m = 20$ for comparison: two-phase with $U_p^t =  7.896763*10^{-31}$ vs single-phase with $U_p^s = 0.000297$. These results suggest that synthetic datasets generated from the two-phase synthesis models achieve a much higher utility than those from the single-phase synthesis model.



\vspace{2mm}
\noindent {\bf Empirical CDF}
We also consider the empirical CDF measure, which assesses the differences between the empirical distribution functions obtained from the original and the synthetic datasets \citep{woo2009global}. Let $O$ denote the original dataset, $S$ denote the synthetic dataset, and $M$ denote the merged dataset. Further, let $D_O$ and $D_S$ be the respective empirical distributions. We can then calculate two empirical CDF measures, $U_m$ and $U_s$:
\begin{eqnarray}
      U_m = \textrm{max}_{1 \le i \le 2n}|D_O(M_i) - D_S(M_i)|,\\
      U_s = \frac{1}{2n}\sum_{i=1}^{2n}[D_O(M_i) - D_S(M_i)]^2,
\end{eqnarray}
where $U_m$ is the maximum absolute difference and $U_s$ is the averaged squared difference between the empirical CDFs. For both quantities, lower values indicate higher utility. 


We calculate the two empirical CDF measures of each of the $m = 20$ merged datasets, for the two-phase and the one-phase respectively. We report the average across the $m = 20$ for comparison: two-phase with $U_m^t = 0.10325$ and $U_s^t = 0.002740404$ vs single-phase with $U_m^s = 0.24164$ and $U_s^s = 0.01870444$. These results once again suggest that synthetic datasets generated from the two-phase synthesis models achieve a higher utility than those from the single-phase synthesis model.


In summary, the two-phase synthesis models produce synthetic datasets with higher global utility. Next, we look at several analysis-specific measures, including density plots of the marginal distribution, the mean, several extreme quantiles, and some regression coefficient estimates of the synthesized income values.

\begin{figure}[h]
    \centering
    \includegraphics[scale=0.4]{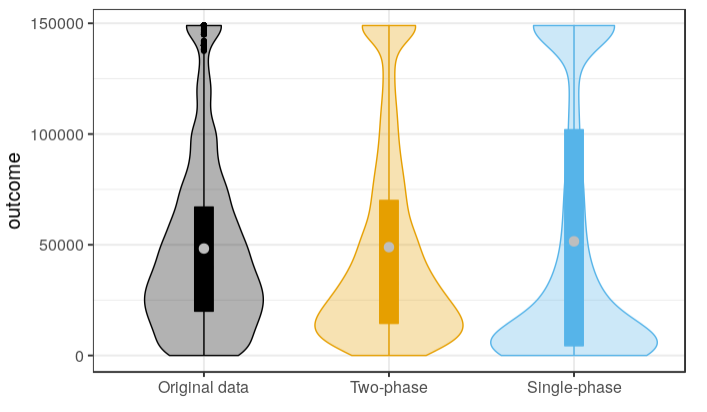}
    \caption{Violin plot of original, two-phase synthetic, and single-phase synthetic income.}
    \label{fig:violin}
\end{figure}

\vspace{2mm}
\noindent {\bf Marginal distribution}
Figure~\ref{fig:violin} depicts the violin plots (with embedded boxplots) of the original income, the synthetic income from the two-phase, and the synthetic income from the single-phase (a randomly selected synthetic dataset out of $m = 20$ synthetic datasets is plotted). We observe that the two-phase synthesis models have performed much better at preserving the marginal distribution of income than the single-phase synthesis model in various aspects: its mean estimate (the grey dot) is closer to the true mean; it preserves the general shape of the income distribution; its interquartile range (IQR) expressed by the embedded boxplot is much closer to that of the original data; the portions in the tails are better preserved, albeit there are more synthesized records with extremely low and high income values compared to the original data.

\vspace{2mm}
\noindent {\bf Mean and extreme quantiles}
To further evaluate analysis-specific utility, we look at the mean estimate and several extreme quantile estimates of the synthesized income values. For the mean estimate, we use the combining rules of partially synthetic data \citep{Drechsler2011book}; the details are included in the Appendix for brevity. For the quantile estimates, we use the bootstrapping method.

We obtain the point estimate and 95\% confidence interval estimate of each quantity of interest. Moreover, we calculate the interval overlap measure to compare the closeness between confidence intervals \citep{Karr2006}:
\begin{eqnarray}
      I = \frac{U_i - L_i}{2(U_o - L_o)} + \frac{U_i - L_i}{2(U_s - L_s)},
\end{eqnarray}
where $[L_o, U_o]$ denotes the confidence interval constructed from the original dataset, and $[L_s, U_s]$ denotes the average of confidence intervals constructed from the $m = 20$ synthetic datasets. Moreover, $L_i = \textrm{max}(L_o, L_s)$, and $U_i = \textrm{min}(U_o,U_s)$. On the one hand, the closer-to-1 the interval overlap measure $I$ is, the higher the overlap between two intervals, indicating higher utility. On the other hand, the further away from 1 $I$ is, the lower the overlap between two intervals, indicating lower utility. $I$ can be negative, and the absolute value of a negative $I$ increases as two intervals are further away.

\begin{table}[H]
    \centering
    \begin{tabular}{p{1in} p{1in} p{1.5in} p{1in}}
    \hline
    Dataset  & Point estimate & $95\%$ C.I. & Interval overlap \\
    \hline
    Original  & 48289.12 & [47206.82, 49371.42] & \\
    Two-phase & 48734.40 &  [47445.45, 50023.35] &  0.8184317\\
    Single-phase & 52884.21 & [51198.93, 54569.49] &  -0.6932335\\
     \hline
    \end{tabular}
    \vspace{1mm}
    \caption{Point estimate, 95\% C.I., and interval overlap of mean income.}
    \label{tab:mean}
\end{table}

Results and comparisons of the mean estimate are in Table~\ref{tab:mean}. The point estimate from the two-phase synthesis models is much closer to that from the original data, with a high level of interval overlap of approximately 0.82. By contrast, the point estimate from the single-phase synthesis model is far off, with a non-overlapping interval compared to that from the original dataset.

Results and comparisons of two extreme quantile estimates, the 10-th and the 80-th, are in Tables~\ref{tab:quantile10} and \ref{tab:quantile80}, respectively. The two-phase synthesis has preserved the 10-th quantile relatively well with a high interval overlap measure of approximately 0.70, while that of the single-phase synthesis is negative with a large absolute value of approximately 7.71. However, for the 80-th quantile, neither of the two syntheses has produced an overlapping interval, though clearly the two-phase is much better than the single-phase. These results resonate with our observation in Figure \ref{fig:violin}: extreme quantiles are not preserved as well as some estimates, such as the mean. Nevertheless, the two-phase synthesis has done a better job preserving extreme quantiles, especially with point estimates: the two-phase synthesis estimates are much closer to the original estimates when compared to the single-phase synthesis point estimates. Results of other extreme quantiles reach the same conclusion.


\begin{table}[h]
    \centering
    \begin{tabular}{p{1in} p{1in} p{1.5in} p{1in}}
    \hline
    Dataset & Point estimate & $95\%$ C.I. & Interval overlap \\
    \hline
     Original  & 6132.53 & [5000.00, 7400.70] & \\
    Two-phase & 6014.41 & [5547.33, 6502.99] & 0.6990378\\
    Single-phase & 1169.26  & [1039.03, 1305.37] &  -7.705418\\
     \hline
  \end{tabular}
  \vspace{1mm}
    \caption{Point estimate, 95\% C.I., and interval overlap of the 10-th quantile of income.}
    \label{tab:quantile10}
\end{table}

 \begin{table}[h]
    \centering
    \begin{tabular}{p{1in} p{1in} p{1.5in} p{1in}}
    \hline
    Dataset &  Point estimate & $95\%$ C.I. & Interval overlap \\
    \hline
    Original  & 75285.72  & [74000.00, 78000.00] & \\
    Two-phase & 84420.57 & [80594.71, 88246.27] & -0.4938931\\
    Single-phase & 147674.80 & [142567.80, 149000.00] &  -13.09008\\
     \hline
    \end{tabular}
    \vspace{1mm}
    \caption{Point estimate, 95\% C.I., and interval overlap of the 80-th quantile of income.}
    \label{tab:quantile80}
\end{table}

\vspace{2mm}
\noindent {\bf Regression coefficients}
Our final set of evaluation of analysis-specific utility is regression coefficients from a linear regression model of the continuous income regressed on all available predictors. The same combining rules and interval overlap measure are used. Results and comparisons of the Age regression coefficient are in Table~\ref{tab:regression_age}. Once again, the point estimate and interval overlap results from the two-phase synthesis models indicate a higher level of utility preservation compared to the single-phase synthesis model. We recognize that due to the skewedness of income, data analysts might conduct analyses other than this regression model. 


\begin{table}[h]
    \centering
    \begin{tabular}{p{1in} p{1in} p{1.5in} p{1in}}
    \hline
    Dataset & Point estimate & $95\%$ C.I. & Interval overlap \\
    \hline
    Original  & 394.09 & [331.08, 457.11] & \\
    Two-phase & 403.38  & [321.14, 485.61]   & 0.8831511\\
    Single-phase &  323.67 & [208.66, 438.67] & 0.6607438 \\ 
    \hline
    \end{tabular}
    \vspace{1mm}
    \caption{Point estimate, 95\% C.I., and interval overlap of the Age regression coefficient.}
    \label{tab:regression_age}
\end{table}
    

In summary, we have considered a wide range of analysis-specific utility measures. All results suggest a higher level of utility preservation of the two-phase synthesis models. Together with its superior performance of global utility preservation, we conclude that the two-phase synthesis models produce synthetic data with high data utility. The two-phase design adequately captures the distributional characteristics of the original data.

\subsection{Risk evaluations: measures and results}
\label{results:risk}

\vspace{2mm}
\noindent {\bf Identification disclosure risk} Identification disclosure risk refers to the risk of an intruder correctly identifying targeted records in the synthetic datasets, and subsequently learning sensitive information about the identified target records \citep{hu2018bayesian}. As income is the only synthesized variable, we assume that the intruder knows the true income value and tries to identify the person in the released data.

We follow the identification risk evaluation approach from \citet{ReiterMitra2009}. Let $c_i$ be the number of records with the highest match probability for target record $i$ (records sharing same / similar known variables \emph{and} same / similar synthesized variables). Let $T_i = 1$ if the true match is among $c_i$, and $T_i  = 0$ otherwise. Additionally, let $K_i = 1$ if $c_iT_i = 1$ (if true match is unique), and $K_i = 0$ otherwise. Similarly, let $F_i = 1$ if $c_i(1-T_i) = 1$ (if there exists unique match but it is not true match), and $F_i = 0$ otherwise. Finally, let $s$ be the number of uniquely-matched records (i.e. $\sum_{i=1}^{n}\mathbb{I}(c_i = 1)$, where $\mathbb{I}(\cdot)$ is a binary indicator).


The three measures we consider are the expected match risk, the true match rate, and the false match rate \citep{ReiterMitra2009}. The expected match risk quantifies the average probability of performing a correct identity match for any record $i$:
\begin{eqnarray}
    E = \frac{1}{n}\sum_{i=1}^{n}\frac{T_i}{c_i},
\end{eqnarray}
where $n$ is the number of records. A higher expected match risk indicates higher identification risk. 
The true match rate refers to the percentage of true and unique matches, quantified as:
\begin{equation}
    T = \sum_{i=1}^{n}\frac{K_i}{n}.
\end{equation}
$T \in [0, 1]$, and a higher true match rate indicates higher identification risk, and vice versa.
The false match rate refers to the percentage of unique matches that are false matches:
\begin{equation}
    F = \sum_{i=1}^{n}\frac{F_i}{s}.
\end{equation}
$F \in [0, 1]$, and unlike the first two measures, a higher false match rate indicates lower identification risk~\citep{hu2018bayesian}.

To calculate $c_i$, the number of records with the highest match probability for target record $i$ for a continuous synthesized variable, such as Income in our NHIS application, we assume that the intruder knows three out of the seven available un-synthesized variables of this record: Sex, Race, and Education. We further assume that for target record $i$, the intruder knows the true value of the income. With such information, the intruder attempts to identify record $i$ in the synthetic datasets. 

We calculate $c_i$ by first searching for all the records in a synthetic dataset sharing the same known variable information as record $i$. Among these records, we find the records whose synthetic income is close to the true income of record $i$. We define ``close" by creating an interval of radius $r$ from record $i$'s true income value. We use a percentage radius $r$ to reflect the magnitude of each record's income value \citep{respp_arxiv}.


\begin{table}[]
    \centering
    \begin{tabular}{p{1.5in}  p{1in} p{1in}}
    \hline
     Risk measure  & Two-phase & Single-phase\\
    \hline
    Expected match risk & 0.00002137 &  0.00001750 \\
    True match rate  & 0 & 0\\
    False match rate & 1 & 1\\
    \hline
    \end{tabular}
    \vspace{1mm}
    \caption{Risk measures for two-phase and single-phase income synthesis.}
    \label{tab:risk}
\end{table}

Table~\ref{tab:risk} presents the three risk measures based on a $r = 0.3$ radius for both two-phase and single-phase income syntheses. The expected match risk measure indicates that the two-phase synthesis produces synthetic data with slightly higher identification risk, compared to the single-phase synthesis: two-phase synthesis produces a higher expected match risk, but the same true match rate and false match rate. We also examine other choices of $r$ and reach the same conclusion.

\vspace{2mm}
\noindent {\bf Attribute disclosure risk} We also consider the attribute disclosure risk. Attribute disclosure happens when the intruder correctly infers the true values of synthesized variables in the publicly released synthetic datasets \citep{hu2018bayesian}. We follow the methods described in ~\cite{reiter2014bayesian} and ~\cite{hu2014disclosure}, and calculate the probability of correctly inferring the true value income of each record. Specifically, we consider 10 records in the neighborhood of record $i$; each candidate record is different from record $i$ only in the income value. Therefore, the calculated probability is the normalized attribute disclosure probability of the true value among 11 records (10 neighborhood records plus the record itself). The higher the probability is, the higher the attribute disclosure risk for a given record. The attribute risk computation is based on a randomly selected synthetic dataset out of $m$ synthetic datasets.

\begin{figure}[h]
    \centering
    \includegraphics[scale=0.35]{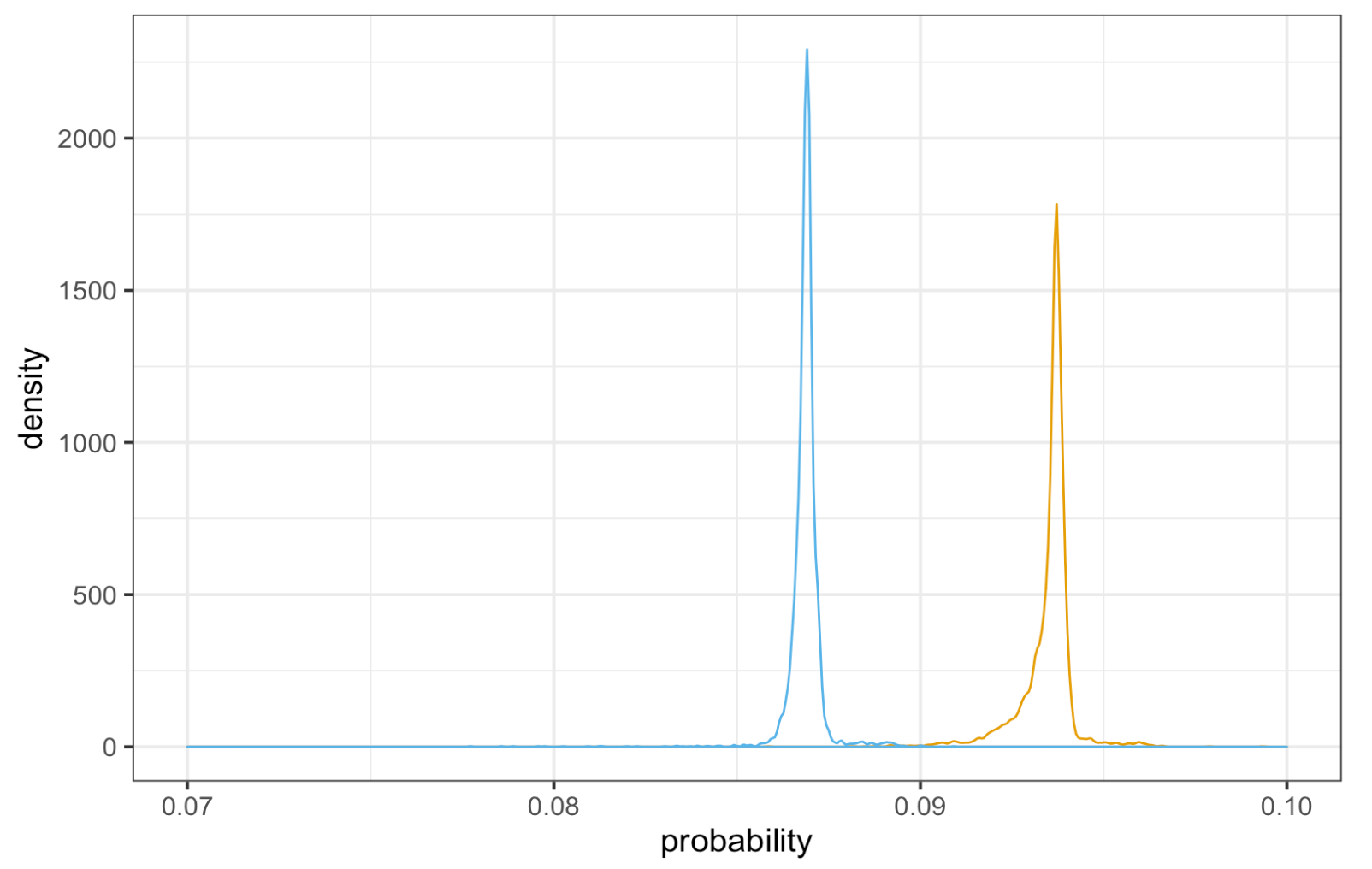}
    \caption{Density plot of probability of two-phase synthesis (orange) and single-phase synthesis (blue).}
    \label{fig:density}
\end{figure}

A density plot of these calculated probabilities is presented for the two-phase synthesis and single-phase synthesis in Figure~\ref{fig:density}. The probability of the true income value being guessed correctly is slightly higher in the two-phase synthesis compared to the single-phase synthesis. We also report the ranking of the probability among the 11 records: the closer the ranking is to 1, the higher the attribute disclosure risk, and vise versa. Results show that the two-phase synthesis has higher attribute disclosure risks, and we include a histogram of ranking for the two-phase and single-phase in the Appendix.

In summary, the level of utility preservation of the two-phase synthesis is overall very high, and much higher than that of the single-phase synthesis. The higher level of utility comes at a price of slightly higher identification and attribute disclosure risks, which is known as the utility-risk trade-off, a well-demonstrated phenomenom in many applications \citep{DuncanStokes2012CHANCE, DrechslerHu2018, respp_arxiv}.

\section{Concluding Remarks}
\label{discussion}
Due to the wide availability of socioeconomic, demographic, and health characteristics, the risk for intruders to use the NHIS to derive confidential information is a concern. Our proposed two-phase synthesis process is a practical method for synthesizing income using Bayesian logistic and linear regressions. We found that a two-phase approach is more effective at preserving the relationships of the variables while maintaining a satisfactory level of disclosure risks when compared to a single-phase approach. 

Our proposed two-phase synthesis process is general and applies to non-Bayesian synthesis models, such as CART synthesis models. In fact, our proposed two-phase synthesis approach can be applied to any dataset containing a sensitive continuous variable with various distributional characteristics to be protected. It is designed to tackle the challenge of preserving records with value of 0 in the synthetic data, an obvious drawback of any single-phase synthesis model.




Specifically for the NHIS data sample, further exploration of additional predictor variables is one future work direction. For example, variables related to medical care access, health behaviors, occupation, or family interrelationships might improve the utility preservation level of simulated synthetic data, especially on the extreme quantile estimates.



\bibliographystyle{natbib}
\bibliography{synbib}

\newpage
\section*{Appendix}

\subsection*{1. Data cleaning steps of the NHIS data sample in Section \ref{intro:data}}


72,831 observations were collected by the U.S. Census Bureau. Specifically, the health and healthcare access information for this study was drawn from the NHIS. 

Our data cleaning process includes multiple steps. First, all missing and not available (NA) observations were removed. Next, NIU (Not In Universe) values, expressed as 0 and 00, were deleted from Education, HoursWorked, HealthInsurance, and HomeOwnership. For Education, HoursWorked, and HomeOwnership, all rows that contained a variable value of 97 (Refused), 98 (Unknown- not ascertained), and 99 (Unknown - don't know) were removed. Similarly, this was done with HealthInsurance for values of 7, 8, and 9, as well as for Race with values 970, 980, and 990. This reduced the sample size from 72,831 to 33,599 observations. Because of computational limitations, we conducted our investigation using a random sample of 5000 entries. 

The variable Race was re-coded into 5 main racial backgrounds demonstrated in Table~\ref{tab:variable}. Education was expressed by education attainment completed by grade, which was collapsed to 3 categories of 4 years of high school or less, 4 years of college, and 5+ years of college. Thus, each category is representative of a greater sample size.

\subsection*{2. Combining rules for partially synthetic data}

\begin{eqnarray}
      \bar{q}_m = \sum_{\ell=1}^{m}\frac{q^{(\ell)}}{m},\\
      b_m = \sum_{\ell=1}^{m}\frac{(q^{(\ell)} - \bar{q}_m)^2}{m-1},\\
      \bar{v}_m = \sum_{\ell=1}^{m}\frac{v^{(\ell)}}{m},
\end{eqnarray}
where $q^{(\ell)}$ and $v^{(\ell)}$ are the point and variance estimates for the income in each synthetically generated dataset $\ell$. 

Let $\bar{q}_{m}$ be the point estimate of the partially synthetic dataset. The variance estimate of $\bar{q}_{m}$ is $T_{p}$, which can be express as 
\begin{eqnarray}
        T_{p} = \frac{b_{m}}{m} + \bar{v}_{m}.
\end{eqnarray}

Next, in order to make inferences, use a $t$ distribution with degrees of freedom 
\begin{eqnarray}
        v_{p} = (m - 1) \left(1 + \frac{\bar{v}_{m}}{\frac{b_{m}}{m}}\right).
\end{eqnarray}

To ultimately obtain the 95\% confidence interval estimate, we have 
\begin{eqnarray}
        \left(\bar{q}_{m} - t_{v_{p}}(0.975) \times \sqrt{\frac{b_{m}}{m} + \bar{v}_{m}},\quad  \bar{q}_{m} + t_{v_{p}}(0.975) \times \sqrt{\frac{b_{m}}{m} + \bar{v}_{m}}\right),
\end{eqnarray}
where $t_{v_{p}}$ is the $t$ score at 0.975 with degrees of freedom $v_{p}$.

\subsection*{3. Histogram of ranking in Section \ref{results:risk}}

\begin{figure}[h]
    \centering
    \includegraphics[scale=0.4]{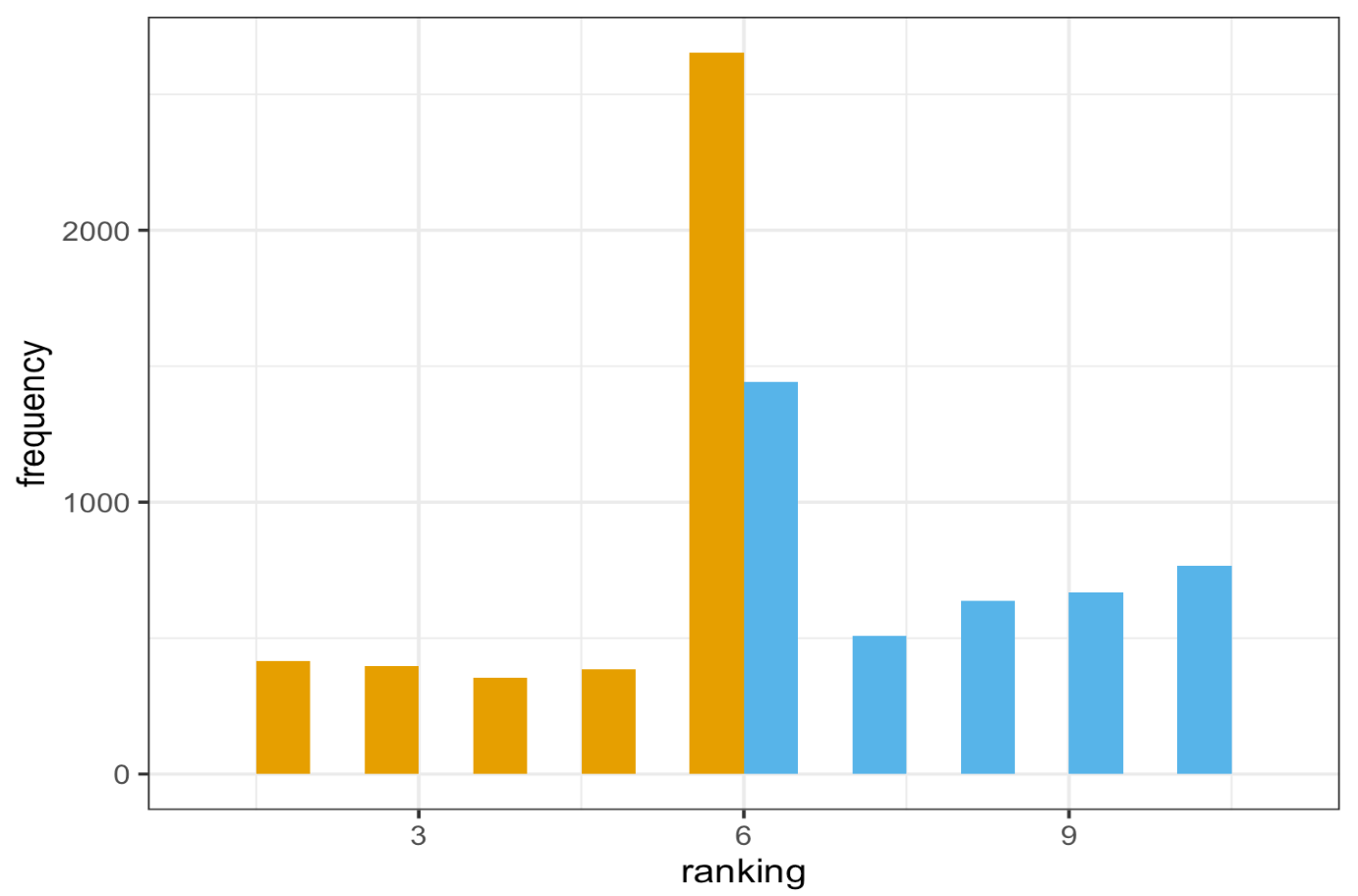}
    \caption{Histogram of ranking of two-phase synthesis (orange) and single-phase synthesis (blue).}
    \label{fig:histogram}
\end{figure}

\end{document}